\documentstyle[12pt]{article}

\newcommand{\orm}{{\rm o}}
\newcommand{\intrm}{{\rm int}}
\newcommand{\extrm}{{\rm ext}}
\newcommand{\tanrm}{{\rm tan}}

\newcommand{\ug}{\; = \;}

\newcommand{\equ}{\; \equiv \;}
\newcommand{\Vbf}{\mbox{\boldmath $V$}}
\newcommand{\jbf}{\mbox{\boldmath $j$}}

\newcommand{\Ebf}{\mbox{\boldmath $E$}}

\newcommand{\Abf}{\mbox{\boldmath $A$}}
\newcommand{\ebf}{\mbox{\boldmath $e$}}

\newcommand{\Hbf}{\mbox{\boldmath $H$}}

\newcommand{\nbf}{\mbox{\boldmath $n$}}
\newcommand{\hatnbf}{{\hat{\nbf}}}

\newcommand{\nabf}{\mbox{\boldmath $\nabla$}}

\newcommand{\pa}{\partial}

\newcommand{\text}{\rm}

\newcommand{\drm}{{\rm d}}
\newcommand{\grm}{{\rm g}}

\newcommand{\Ccal}{{\cal C}}
\newcommand{\ga}{\gamma}

\newcommand{\infi}{\infty}

\newcommand{\ra}{\rightarrow}

\newcommand{\ze}{\zeta}

\newcommand{\bb}{\begin{equation}}
\newcommand{\ee}{\end{equation}}
\newcommand{\bega}{\begin{eqnarray}}
\newcommand{\ega}{\end{eqnarray}}
\newcommand{\begae}{\begin{eqnarray*}}
\newcommand{\egae}{\end{eqnarray*}}

\newcommand{\h}{\hspace*{4ex}}
\newcommand{\dis}{\displaystyle}

\newcommand{\Om}{\Omega}
\newcommand{\om}{\omega}

\newcommand{\cent}{\centerline}
\newcommand{\vs}{\vspace*}

\begin{document}

\baselineskip 0.65cm

\begin{center}

{\large {\bf The X-shaped, localized field generated by a Superluminal
electric charge}$^{\: (\dag)}$}
\footnotetext{$^{\: (\dag)}$
Work partially supported by INFN and MURST/MIUR (Italy), and by FAPESP (Brazil).
\ E-mail address for contacts: recami@mi.infn.it }

\end{center}

\vs{5mm}

\centerline{ Erasmo Recami,$^{{\rm a},{\rm b}}$ Michel
Zamboni-Rached,$^{\rm c}$ C\'esar A.Dartora,$^{\rm c}$ and K. Z.
N\'obrega.$^{\rm c}$ }

\vs{0.2 cm}

\cent{$^{\rm a}${\em Facolt\`a di Ingegneria, Universit\`a statale di Bergamo,
Dalmine (BG), Italy;}}
\cent{$^{\rm b}${\em INFN---Sezione di Milano, Milan, Italy; \ {\rm and}}}
\cent{$^{\rm c}${\em D.M.O., FEEC, State University at Campinas,
Campinas, SP, Brazil.}}

\vs{1.5 cm}

{\bf Abstract  \ --} \ It is now wellknown that Maxwell equations admit of
wavelet-type solutions endowed with arbitrary group-velocities ($0 <
v_\grm < \infi$). \ Some of them, which are rigidly moving and have been
called localized solutions, attracted large attention. \ In particular,
much work has been done with regard to the Superluminal localized
solutions (SLS), the most interesting of which resulted to be the ``X-shaped"
ones. The SLSs have been actually produced in a number of experiments,
always by suitable interference of ordinary-speed waves. \ In this note we
show, by contrast, that even a Superluminal charge creates an electromagnetic
X-shaped wave. Namely, on the basis of Maxwell equations, we are able to
evaluate the field associated with a Superluminal charge (under the
approximation of pointlikeness): it results to constitute a very simple
example of true X-wave.\\

\



\newpage

{\bf 1. -- Introduction}\\

\h It is well-known that Maxwell equations have been shown to admit of
wavelet-type solutions endowed with arbitrary[1] group-velocities $0 <
v_\grm < \infi$. \ Some of them, which are rigidly moving and have been called
``localized solutions", attracted large attention[2]. \ In particular,
much work has been done with regard to the Superluminal localized
solutions (SLS), the most interesting of which ---as predicted by
Special Relativity (SR) itself[3]--- resulted to be the ``X-shaped" ones[4]. \
Such X-shaped SLSs have been actually produced in a number of
experiments[5].

\h The theory of SR, when based on the {\em ordinary} postulates but not
restricted to subluminal waves and objects, i.e., in its extended version[6],
predicts the simplest X-shaped wave to be the one corresponding to the
electromagnetic field created by a
Superluminal\footnote{Incidentally, let us recall that the {\em luminal}
case was successfully examined by Bonnor[7], who showed the Maxwell equations
to admit of finite-energy solutions even in the limiting case of a
(mass-free) ``particle" carrying equal amounts of positive and negative
electric charge.}  charge[8]. \ It is of the utmost importance
evaluating the field associated with a Superluminal electric
charge\footnote{For simplicity, we are here adopting the standard language,
but it should be recalled that we must rather think in terms of an
``electromagnetic charge", which behaves as an {\em electric} charge
when subluminal and as a {\em magnetic} pole when Superluminal: cf.
refs.[9].}, not only as a contribution to the theory of X-shaped waves,
but also as a starting point for predicting the
electromagnetic interaction of a charged ``tachyon" with ordinary matter
(and planning, maybe, the construction of a suitable detector).\\

\

{\bf 2. -- The toy-model of a pointlike Superluminal charge}\\
 
\h Let us first start by considering, formally, a pointlike Superluminal
charge, even if the hypothesis of pointlikeness (already unacceptable
in the subluminal case) is totally inadequate in the Superluminal
case, as it was thoroughly shown in refs.[8].

\h Then, let us consider the ordinary vector-potential $A^\mu$ and a
current desity $j^\mu \equiv (0,0,j_z;j^\orm)$ flowing in the $z$-direction.
On assuming the fields to be generated by the sources only, one has that
$A^\mu \equiv (0,0,A_z;\phi)$, which, when adopting the Lorentz gauge, obeys
the equation $\Box A^\mu = j^\mu$. \ Such non-homogeneous wave equation,
in cylindrical co-ordinates $(\rho,\theta,z;t)$ and for axial symmetry [which
requires a priori that $A^\mu = A^\mu(\rho,z;t)$], 
writes:\footnote{As a further check of our calculations, we started also
from the so-called scalar Bromwich--Borgnis[10] potential $u$, under the
hypothesis that $\jbf = (0,0,j_z)$, in which case it is \ $E_\rho =
\pa^2 u / \pa\rho \, \pa z$; \ while \ $E_z = -\pa^2 u / {\pa\tau}^2 + \pa^2 u /
{\pa z}^2$; \ and \ $B_\phi = \pa^2 u / \pa\rho \, \pa\tau$, \ where $\tau =
ct$. \ On defining the function $\psi \equiv A_z \equiv {\pa u / \pa \tau}$,
we showed by Maxwell equations that $\psi$ has to obey the same non-homogeneous
(axially symmetric) wave equation (1), with $\mu=3$.} 

\

\hfill{$
\dis{{{\pa^2 A^\mu} \over {\pa\tau^2}} - {1 \over \rho} \; {\pa \over
{\pa\rho}} \, \left(\rho {{\pa A^\mu} \over {\pa\rho}}\right) - {{\pa^2 A^\mu}
\over {\pa z^2}} \ug j^\mu} \ ,
$\hfill} (1)

\

where $\tau \equiv ct$, and $\mu$ assumes the two values $\mu = 3,0$ only.

\h We shall now choose the ``$V$-cone variables"[11], with \ $V^2 > c^2$,

\

\hfill{$
\left\{\begin{array}{clr}
\ze \equ z-Vt  \\
\eta \equ z+Vt
\end{array} \right.
$\hfill} (2)

\

and rewrite eq.(1) as

\

\hfill{$
\dis{\left[-\rho {\pa \over {\pa\rho}} \left(\rho {\pa \over {\pa\rho}}\right)
+ \frac{1}{\ga^2} \frac{\pa^2}{\pa \ze^2} + \frac{1}{\ga^2} \frac{\pa^2}{\pa \eta^2}
- 4 \frac{\pa^2}{\pa\ze \pa\eta} \right] \; A^\mu(\rho,\ze,\eta) \ug
j^\mu(\rho,\ze,\eta)} \ ,
$\hfill} (3)

\

where[6] \ $A^\mu \equiv (0,0,A_z;\phi)$ \ and

\

\hfill{$
\dis{\ga^2 = \frac{1}{\frac{V^2}{c^2} -1}} \ .
$\hfill} (3')

\

\h Let us now suppose $A^\mu$ to be actually independent of $\eta$, namely,
$A^\mu = A^\mu (\rho, \ze)$. \ Due to eq.(3), we shall have $j^\mu =
j^\mu (\rho, \ze)$ too; and therefore $j_z = V j^0$ (from the continuity
equation), and $A_z = V \phi / c$ (from the Lorentz gauge). \ Then, by
calling

$$\psi \equiv A_z \ ,$$

eq.(3) yields the two equations

\

\hfill{$
\dis{\left[- \frac{1}{\rho}\frac{\pa}{\pa\rho} \left(\rho \frac{\pa}{\pa\rho}\right)
+ \frac{1}{\ga^2} \frac{\pa^2}{\pa\ze^2}\right] \; \psi(\rho,\ze)
\ug j_z(\rho,\ze)}
$\hfill} (4a)

\

and

\

\hfill{$
\dis{\phi \ug {c \over V} \, \psi}
$\hfill} (4b)

\

One can notice that the procedure leading to eqs.(4) constitutes a simple
{\em generalization} of Lu et al.'s theorem[12] for non-homogeneous equations,
i.e., for the case with sources.[13]

\h As announced above, let us finally analyse the possibility and
consequences of having a Superluminal pointlike charge, $e$, traveling with
constant speed  $V$ along the $z$-axis ($\rho = 0$) in the positive direction:

\

\hfill{$
\dis{j_z \ug e \, V \, \frac{\delta(\rho)}{\rho} \, \delta(\ze)} \ .
$\hfill} (5)
 
\

Equation (4a) becomes, then, the hyperbolic equation

\

\hfill{$
\dis{\left[-\frac{1}{\rho}\frac{\pa}{\pa\rho}\left(\rho \frac{\pa}{\pa\rho}\right)
+ \frac{1}{\ga^2} \frac{\pa^2}{\pa\ze^2}\right] \, \psi \ug e V \,
\frac{\delta(\rho )}{\rho} \, \delta(\ze)} \ .
$\hfill} (6)

\

To solve it, let us apply (with respect to the variable $\rho$) a
Fourier-Bessel (FB) transformation, carrying by definition a function $f(x)$
into the function $F(\Om)$ as follows

\

\hfill{$
f(x) \ug \dis{\int_0^{\infty} \Om f(\Om) J_0(\Om x) \, \drm\Om} 
$\hfill} (7)

\

\hfill{$
f(\Om ) \ug \dis{\int_0^{\infty} x f(x) J_0(\Om x)\, \drm x} \ ,
$\hfill} (7')

\

$J_0(\Om x)$ being the ordinary zero-order Bessel function. \ Equation (6)
gets transformed, after some calculations, into 

\

\hfill{$
\dis{\left[\frac{1}{\ga^2} \frac{\pa^2}{\pa\ze^2} + \Om^2 \right] \,
\Psi(\Om,\ze) \ug e V \, \delta(\ze)} \ .
$\hfill} (8)

\

By applying subsequently the ordinary Fourier transformation with respect to
the variable $\ze$ (going on, from $\ze$, to the variable $\om$), after
some further manipulations we obtain

\

\hfill{$
\dis{\Psi(\Om,\om) \ug e V \, \frac{\ga^2}{(\ga^2 \Om^2 - \om^2)}} \ .
$\hfill} (9)

\

\h Finally, the solution of eq.(6) is got by performing the corresponding {\em
inverse} Fourier and FB transformations:

\

\hfill{$
\dis{\psi(\rho,\ze) \ug e V \ga^2 \int_{-\infty}^{\infty} \, \drm\om \,
\int_0^{\infty} \, \drm\Om \; \frac{\Om \, J_0(\Om\rho) e^{-i\om\ze}}
{(\ga^2 \Om^2 - \om^2)}} \ ,
$\hfill} (10)

\

which, on using formulae (3.723.9) and (6.671.7) of ref.[14], yields\footnote{In the following we shall put $c = 1$,
whenever convenient.} [$\ze \equiv z - Vt$]:

\

\hfill{$
\left\{\begin{array}{clr}

\psi(\rho,\ze) \ug 0 \ \ \ \ \ \ \ \ \ \ \ \ \ \ \ \ \ \ \ \ \ \ \ \ \ \ \ \ \ \
{\rm {for}} \ \ \ 0 < \ga\mid{\ze}\mid < \rho \\

\\

\dis{\psi(\rho,\ze) \ug \dis{e \frac{V}{\sqrt{\ze^2 - \rho^2(V^2-1)}}}}
\ \ \ \ \ {\rm {for}} \ \ \ 0 \le \rho < \ga\mid{\ze}\mid \ .

\end{array} \right.
$\hfill} (11)

\

In Fig.1 we show our solution $A_z \equiv \psi$, as a function of $\rho$
and $\ze$, evaluated for $\ga = 1$ (i.e., for $V = c \sqrt{2}$). Of course,
we skipped the points in which $A_z$ must diverge, namely the vertex and
the cone surface.

\h For comparison, one may recall that the {\em classical} X-shaped
solution[4] of the {\em homogeneous} wave-equation ---which is shown, e.g.,
in Fig.2--- has the form[11] (with $a > 0$):

\

\hfill{$
\dis{X \ug {\frac{V}{\sqrt{(a-i\ze)^2 + \rho^2(V^2-1)}}}}
$\hfill} (12)

\

In the second one of eqs.(11) it enters expression (12) with the spectral
parameter[11] $a = 0$, which indeed corresponds to the non-homogeneous
case [the fact that for $a=0$ these equations differ for an imaginary unit
will be discussed elsewhere].                    

\h It is quite important, at this point, to notice that {\em such a solution,}
(11), {\em represents a wave existing only inside the} (unlimited) {\em double
cone\/} $\Ccal$ generated by the rotation around the $z$-axis of the straight
lines $\rho = \pm \ga\ze$: \ This is in full agreement with the predictions[15]
of the ``extended" theory of Special Relativity[6].\\

\

{\bf 3. -- Evaluating the fields generated by the Superluminal charge}\\
 
\h Once the solution (11) for the ``potential" $\psi$ has been found, we
can evaluate the corresponding electromagnetic fields. The standard
relations $\Ebf = -\nabf \phi - \pa\Abf / \pa t$ and $\Hbf = \nabf \wedge
\Abf$ imply in the present case [$\psi = \psi(\rho,\ze) \equiv A_z$;
and $\phi = \psi / V$]:

$$\Hbf \ug \dis{-{{\pa\psi} \over {\pa\rho}} \, {\hat{\ebf}}_\phi}$$

$$\Ebf \ug \dis{-{1 \over V} \, {{\pa\psi} \over {\pa\rho}} \, {\hat{\ebf}}_\rho \,
+ \, (V^2 -1) \, {{\pa\psi} \over {\pa\ze}} \, {\hat{\ebf}}_z} \ . $$

Then, from eqs.(11), when $0 \le \rho < \ga\mid{\ze}\mid$ (i.e., inside
the cone $\Ccal$), the fields result to be:\footnote{It should be noticed that
the same results are obtained when starting from the fourpotential associated
with a subluminal charge (e.g., an electric charge at rest), and then
applying to it the suitable Superluminal Lorentz ``transformation"[6].}

\

\hfill{$
E_{\rho} \ug \dis{-\pi e \rho \; \frac{V^2-1}{\sqrt{[\ze^2 - \rho^2(V^2-1)]^3}}}
$\hfill} (13)

\

\hfill{$
E_z \ug \dis{-\pi e \ze \; \frac{V^2-1}{\sqrt{[\ze^2 - \rho^2(V^2-1)]^3}}}
$\hfill} (14)

\

\hfill{$
H_{\phi} \ug \dis{-\pi e V \rho \frac{V^2-1}{\sqrt{[\ze^2 - \rho^2(V^2-1)]^3}}} 
$\hfill} (15)

\

where, let us recall, $\ze \equiv z - Vt$, with $V^2 > c^2$. \  We show in
Fig.3 the direction of the various field components in our co-ordinates; while
the behavior of $E_z$, as a function of $\rho$ and $\ze$, is shown in Fig.4.

\h However, ouside the cone $\Ccal$, i.e., for $0 < \ga\mid{\ze}\mid < \rho$,
one gets as expected that

\

\hfill{$
E_{\rho} \ug E_z \ug H_{\phi} \ug 0 \ .
$\hfill} (16)

\

\h One meets therefore a field discontinuity when crossing the double-cone
surface, since the field is zero outside it. Nevertheless, the boundary
conditions imposed by Maxwell equations[15] are satisfied by our solution (11),
or (13-15), since at each point of the cone surface the electric and the
magnetic field are both tangent to the cone: We shall discuss this point
below. 

\h Here, let us emphasize that, when $V \ra \infi; \ \ga \ra 0$, the electric
field tend to vanish, while the magnetic field tends to the value $H_\phi =
-\pi e / \rho^2$. This does agree with what expected from Extended
Relativity[9], which predicts Superluminal charges to behave, in a sense,
as magnetic monopoles. In the present note we can only mention such a
circumstance, and refer to refs.[3,9]:  Where it is shown that, if one
calls {\em electric} the ``electromagnetic charge" when it is subluminal,
then he should call it {\em magnetic} when Superluminal\footnote{We have shown
elsewhere[16,9] that a Superluminal charge $e$ and a Superluminal current $j^\mu$
are a pseudoscalar and a pseudovector, respectively: like in the case of a
magnetic charge and a magnetic current; so that they should rather be
written as $\ga_5 e$ and $\ga_5 j^\mu$. \ But in this preliminary note
we shall forget about the symmetry properties of those quantities.}
(cf. Fig.46 at page 155 of the first one of refs.[6]). \ Actually, result
(11) can be obtained in a quicker way just by applying
a Superluminal Lorentz ``transformation"[6] to the fields generated by
a subluminal (in particular, at rest) electric point-charge.

\h Let us add, more in general, that ---as mentioned at the end of the
previous Section--- extended relativity predicts, e.g., that the spherical
equipotential surfaces of the electrostatic field created by a charge
at rest get transformed (by a Superluminal Lorentz ``transformation")
into two-sheeted rotation-hyperboloids, contained inside an unlimited
double-cone[6,8]: see Fig.5.  One ought to notice, incidentally, that this
double cone does not have much to do with the Cherenkov cone. In fact the
double cone is associated with a constant-speed Superluminal charge even in
the vacuum, while Cherenkov radiation emission is induced by a fast
electric charge only out of a material medium. \ Moreover (cf. also Fig.27
at page 80 of the first one of refs.[6]) a
Superluminal charge traveling at constant speed, in the vacuum, e.g.,
does {\em not} lose energy[8].

\h We go eventually back to the problem that one meets a field
discontinuity across the double-cone surface [see eqs.(13-15) and
eqs.(16)], since the field is zero outside $\Ccal$; for $\rho \ra \ga
\mid\ze\mid$ the fields (13)-(15) even diverge.  Nevertheless, one
can straightforwardly {\em verify} that our solution (11), or (13-16), satisfies
the following boundary conditions, required by Maxwell equations (in the
present case of a {\em moving} boundary)[17,18]:

\

\hfill{$
\left\{\begin{array}{clr}
\dis{\left(\Ebf_\extrm - \Ebf_\intrm \right) \cdot \hatnbf \ug \sigma}\\

\\

\dis{\left(\Hbf_\extrm - \Hbf_\intrm \right) \cdot \hatnbf \ug 0}\\

\\

\dis{\left(\Ebf_\extrm - \Ebf_\intrm \right)_\tanrm \ug -(\hatnbf \cdot \Vbf)
\, \hatnbf \wedge (\Hbf_\extrm - \Hbf_\intrm)}\\

\\

\dis{\left[1 - (\hatnbf \cdot \Vbf)^2 \right] \, \hatnbf \wedge \left(\Hbf_\extrm
- \Hbf_\intrm \right) \ug \jbf}  \ .

\end{array} \right.
$\hfill} (17)

\

\newpage

{\bf Acknowledgements}\\
The authors are grateful, for useful discussions or kind cooperation, to
V.Abate, C.E.Becchi, M.Brambilla, R.Chiao, C.Cocca, R.Collina, R.Colombi,
G.C.Costa, P.W.Milonni, G.Degli Antoni, F.Fontana,
G.Kurizki, G.Pedrazzini, A.Steimberg, J.W.Swart, M.Villa, and particularly
to H.E.Hern\'andez-Figueroa, R.Mignani, K.Z.N\'obrega, G.Salesi and
A.Shaarawi.

\newpage

FIGURE CAPTIONS\\

\

Fig.1 -- Behaviour of $A_z \equiv \psi$, as a function of $\rho$
and $\ze$, evaluated for $\ga = 1$ (i.e., for $V = c \sqrt{2}$). \ [Of
course, we skipped the points in which $A_z$ must diverge: namely, the
vertex and the cone surface].

\

\

Fig.2 -- Illustration of the real part of a {\em classical} X-shaped wave
(as a function of $\rho$ and $\ze$), evaluated for $V = 5 \, c$ and
$a = 5 \times 10^{-7} \;$m.

\

\

Fig.3 -- The direction is here depicted of the various field components,
in our coordinates.

\

\

Fig.4 -- Behavior of the $z$-component of the electric field generated
by a Superluminal (pointlike) charge as a function of $\rho$ and $\ze$,
with the same parameters used for Fig.1. \ [Once more, we skipped the points
in which $E_z$ must diverge: namely, the vertex and the cone surface].
                                  
\

\

Fig.5 -- The spherical equipotential surfaces of the electrostatic field
created by a charge at rest get transformed into two-sheeted
rotation-hyperboloids, contained inside an unlimited double-cone, when the
charge travels at Superluminal speed (cf. refs.[6,8]). This figures shows,
among the others, that a Superluminal charge traveling at constant speed,
in a homogeneous medium like the vacuum, does {\em not} lose energy[8]. \
Let us mention, incidentally, that this double cone has nothing to do with
the Cherenkov cone (see the text). \ The present picture is a reproduction
of our Fig.27, appeared in 1986 at page 80 of the first one of refs.[6].

\newpage

REFERENCES\\

[1] See, e.g., H.Bateman: {\it Electrical and Optical Wave Motion} (Cambridge
Univ.Press; Cambridge, 1915); \ J.A.Stratton:  {\em Electromagnetic Theory}
(McGraw-Hill; New York, 1941), p.356; R.Courant and D.Hilbert: {\em Methods
of Mathematical Physics} (J.Wiley; New York, 1966), vol.2, p.760. \ See also,
e.g.: V.K.Ignatovich: Found. Phys. 8 (1978) 565; \ J.N.Brittingham: J. Appl. Phys.
54 (1983) 1179; \ R.W.Ziolkowski: J. Math. Phys. 26 (1985) 861; \ J.Durnin,
J.J.Miceli and J.H.Eberly: Phys. Rev. Lett. 58 (1987) 1499; \ P.Hillion:
J. Math. Phys. 29 (1988) 1771; \ A.M.Shaarawi,
I.M.Besieris and R.W.Ziolkowski: J. Math. Phys. 31 (1990) 2511; \
A.O.Barut et al.: Phys. Lett. A143 (1990) 349; Found. Phys. Lett.
3 (1990) 303; Found. Phys. 22 (1992) 1267; Phys. Lett. A180 (1993) 5; A189
(1994) 277; \ P.Hillion: Acta Applicandae Matematicae 30 (1993) 35; \
R.Donnelly and R.W.Ziolkowski: Proc. Roy. Soc. London A440 (1993) 541; \
J.Vaz and W.A.Rodrigues: Adv. Appl. Cliff. Alg. S-7 (1997) 457; \ S.Esposito:
Phys. Lett. A225 (1997) 203.\hfill\break 

[2] Cf., e.g., the brief review-article by E.Recami: Found. Phys. 31 (2001)
1119.\hfill\break

[3] E.Recami et al.: Lett. Nuovo Cim. 28 (1980) 151; 29 (1980) 241; \
A.O.Barut, G. D.Maccarrone and E.Recami: Nuovo Cimento A71 (1982) 509. \
See also E.Recami: refs.[4] and [6];
E.Recami, F.Fontana and R.Garavaglia: Int. J. Mod. Phys. A15 (2000) 2793; \
and E.Recami et al.: Il Nuovo Saggiatore 2(3) (1986) 20; 17(1-2) (2001) 21.
\hfill\break

[4] E.Recami: Physica A252 (1998) 586; \ J.-y.Lu, J.F.Greenleaf
and E.Recami: ``Limited diffraction solutions to Maxwell (and Schroedinger)
equations'', Lanl Archives \# physics/9610012 (Oct.1996); \ R.W.Ziolkowski,
I.M.Besieris and A.M.Shaarawi: J. Opt. Soc. Am. A10 (1993) 75; \
J.-y.Lu and J.F.Greenleaf: IEEE Trans. Ultrason. Ferroelectr. Freq. Control
39 (1992) 19. \ Cf. also M.Zamboni Rached, E.Recami and F.Fontana: Phys. Rev.
E64 (2001) 066603.\hfill\break

[5] J.-y.Lu and J.F.Greenleaf: IEEE Trans. Ultrason. Ferroelectr. Freq.
Control  39 (1992) 441; \ P.Saari and K.Reivelt: Phys. Rev. Lett. 79
(1997) 4135-4138; \ D.Mugnai, A.Ranfagni and R.Ruggeri: Phys. Rev. Lett.
84 (2000) 4830.\hfill\break

[6] See E.Recami: Rivista Nuovo Cim. 9(6) (1986) and refs. therein; cf. also
E.Recami and R.Mignani: Rivista Nuovo Cim. 4 (1974) 209-290, E398.\hfill\break

[7] W.B.Bonnor: Int. J. Theor. Phys. 2 (1969) 373.\hfill\break

[8] See E.Recami: ref.[6], fig.27 and pp.80-81; and refs. therein. \ Cf. also
R.Folman and E.Recami: Found. Phys. Letters 8 (1995) 127-134.\hfill\break

[9] See E.Recami: ref.[6], pp.82-83 and pp.152-156 (in particular, fig.46);
and refs. therein; \ R.Mignani and E.Recami: Lett. Nuovo Cimento 9 (1974)
367-372; Nuovo Cimento A30 (1975) 533-540; Physics Letters B62 (1976)
41-43. \ See also E.Recami (editor): {\em Tachyons, Monopoles, and Related
Topics} (North-Holland; Amsterdam; 1978), pp.X + 285.\hfill\break

[10] T.J Bromwich: Phil. Mag. 38 (1919) 143.\hfill\break

[11] Cf. M.Zamboni Rached, E.Recami and H.E.Harn\'andez-Figueroa: ``New
localized Superluminal solutions to the wave equations with finite total
energies and arbitrary frequencies", Lanl Archives e-print \# physics/0109062,
in press in Europ. Phys. Journ.~D. \ See also M.Zamboni-Rached, K.Z.N\'obrega,
H.E.Hern\'{a}ndez-Figueroa and Erasmo Recami: ``Localized Superluminal
solutions to the wave equation in (vacuum or) dispersive media, for arbitrary
frequencies and with adjustable bandwidth", e-print physics/0209101
(submitted for pub.).\hfill\break

[12] J.-y.Lu, H.-h.Zou and J.F.Greenleaf: IEEE Transactions on 
Ultrasonics, Ferroelectrics and Frequency Control 42 (1995) 850.\hfill\break

[13] For different generalizations, which take the presence of boundaries
into account (as in the cases of cylindrical waveguides, co-axial cables,
etc.), see M.Zamboni Rached, E.Recami and F.Fontana: Phys. Rev. E64 (2001)
066603; \ M.Z.Rached, K.Z.Nobrega, E.Recami \& H.E.Hern\'andez F.: ``Superluminal
X-shaped beams propagating without distortion along a coaxial guide", e-print
physics/0209104, in press in Phys. Rev. E; \ M.Z.Rached, F.Fontana and
E.Recami: ``Superluminal localized solutions to Maxwell equations propagating
along a waveguide: The finite-energy case", e-print physics/0209102,
submitted to Phys. Rev.~E.).\hfill\break

[14] I.S.Gradshteyn and I.M.Ryzhik: {\em Integrals, Series and Products},
4th edition (Ac.Press; New York, 1965).\hfill\break 

[15] See Fig.4 in A.O.Barut, G.D.Maccarrone and E.Recami: Nuovo Cimento A71
(1982) 509, page 518; \ Fig.1 in P.Caldirola, G.D.Maccarrone and E.Recami:
Lett. Nuovo Cim. 29 (1980) 241, page 243; \ and \ E.Recami and G.D.Maccarrone:
Lett. Nuovo Cim. 28 (1980) 151-157. \ Cf. also H.C.Corben: Lett. Nuovo Cim.
11 (1974) 533; Nuovo Cimento A29 (1975) 415.\hfill\break

[16] M.A.Faria-Rosa, E.Recami and W.A.Rodrigues Jr.: Physics Letters B173
(1986) 233-236; B188 (1987) E511; \ A.Maia Jr. E.Recami, W.A.Rodrigues Jr.
and M.A.F.Rosa: J. Math. Phys. 31 (1990) 502-505; Phys.Lett. B220 (1989)
195-199.\hfill\break

[17] Cf., e.g., equations (I.23), (I.24), (I.17) and (I.18) at pages 19--22 of
J.D.Jackson: {\em Classical Electrodynamics}, 2nd edition (J.Wiley; New
York, 1975).\hfill\break

[18] P.D.Noerdlinger: Am. J. Phys. 39 (1971) 191.

\end{document}